\documentstyle[aps,preprint,epsfig]{revtex}

\begin{document}

\title{Phase Transitions in  the Spin-Half $J_1$--$J_2$ Model}

\author{Raymond F. Bishop$^\dagger$, Damian J.J. Farnell$^*$,
and John B. Parkinson$^\dagger$}

\address{$^\dagger$Department of Physics, University of Manchester Institute of
Science and Technology (UMIST), P O Box 88, Manchester M60 1QD, United 
Kingdom.}

\address{$^*$Institut f\"ur Theoretische Physik, Universit\"at zu K\"oln, 
Z\"ulpicher Str., 50674 K\"oln, Germany.}

\maketitle

\abstract{The coupled cluster method (CCM) is a well-known method 
of quantum many-body theory, and in this article we present 
an application of the CCM to the spin-half $J_1$--$J_2$ 
quantum spin model with nearest- and next-nearest-neighbour
interactions on the linear chain and the square lattice. 
We present new results for ground-state expectation values of
such quantities as the energy and the sublattice magnetisation. 
The presence of critical points in the solution of the 
CCM equations, which are associated with phase 
transitions in the real system, is investigated. 
Completely distinct from the investigation of the critical 
points, we also make a link between the expansion coefficients 
of the ground-state wave function in terms of an Ising basis 
and the CCM ket-state correlation coefficients. 
We are thus able to present evidence of the breakdown, 
at a given value of $\frac {J_2}{J_1}$, of the 
Marshall-Peierls sign rule which is known to be satisfied at 
the pure Heisenberg point ($J_2 = 0$) on any bipartite lattice. 
For the square lattice, our best estimates of the points at 
which the sign rule breaks down and at which the 
phase transition from the antiferromagnetic phase to the frustrated
phase occurs are, respectively, given by $\frac {J_2}{J_1} \approx 
0.26$ and  $\frac {J_2}{J_1} \approx 0.61$.}










\section{Introduction}

Antiferromagnetic materials have conveniently been modelled, since the
early work of Heitler and London, as a lattice of magnetic atoms upon
which the active electrons are localised.  Furthermore, the exchange
interactions between the electrons are conventionally described solely in
terms of the spin degrees of freedom of the electrons. An archetypal model
of this class remains the Heisenberg model in which only nearest-neighbour
exchange interactions are included, and these are all taken to be equal.
Although the Heisenberg model on the one-dimensional (1D) chain has been
exactly solved many years ago by Bethe [1], it is still the case that relatively
few other exact solutions have been found in the intervening 65 years or so
to comparable models in higher dimensions or to models involving more
complicated interactions, especially those containing an element of
frustration.

On the other hand, various approximate numerical techniques have by now
been applied to a large number of such magnetic lattice Hamiltonians.  For
example, many variational calculations have been undertaken, employing a
wide variety of trial wave functions. Although these often give accurate upper
bounds for the ground-state energy, for example, one often finds that the 
differences between the estimated energies for trial states of widely differing
kinds are very small.  Hence, predictions based on variational calculations for
properties other than the energy, or to such questions as whether the exact
ground state is ordered or disordered, are notoriously unreliable.

As a common alternative one may perform exact diagonalisations on small
finite clusters of spins drawn from the infinite lattice under consideration.
However, even with modern computers, one considers clusters of
$N$ spins with $N \leq 36$. Extrapolation to the infinite lattice,
$N\rightarrow\infty$, then needs to be performed. While 
exact results from finite-size scaling theory are often of great help in this
regard, the extrapolation does need to be handled with great care. This is
particularly true when using the results from finite clusters to make
quantitative predictions for such quantities as the order parameter. Many
wrong claims have been made in the past from an improper treatment of the
very subtle phenomena which need to be taken into account in the 
extrapolations, as has been stressed and discussed with great care by Lhuillier
and her co-workers [2]. Furthermore, one expects that such finite-cluster
calculations will become less accurate the closer one approaches a quantum
phase transition between states of different quantum order, marked by a 
critical value of some coupling parameter, at which a correlation length
characterising the order typically diverges.

Results for much larger clusters are typically obtained by stochastic
simulation of the many-body Schr\"{o}dinger equation using various
quantum Monte Carlo (QMC) algorithms. Where the basic spin-lattice
Hamiltonian can be mapped onto an equivalent bosonic problem, as in the 
case of the Heisenberg model on a bipartite lattice, such QMC techniques
can readily be applied to clusters containing several hundred or more spins,
and very accurate results thereby obtained. In these cases, such as the 
Heisenberg model on the two-dimensional (2D) square lattice [3,4], the QMC
results can usually be considered as ``exact'', with the resulting errors 
arising only, or largely, from statistical errors which are open to systematic
reduction within the limits of the available computing power.

What ultimately underpins these bosonic mappings, and what therefore makes
such QMC simulations so readily attainable, is the knowledge that in some
appropriate representation the multi-spin cluster coefficients describing the 
{\it N}-body wave function are all positive-definite. 
For example, in the case of the
Heisenberg model on a bipartite lattice, this information is provided by the 
well-known Marshall-Peierls sign rule theorem [5].

Conversely, where such prior knowledge of the nodal structure of a many-body
wave function is not exactly known, QMC calculations are beset by the 
notorious ``minus-sign problem'', and are usually then much less reliable
or much more difficult to implement with known algorithms. A typical way
that such complications arise in spin-lattice problems is from the introduction
of frustration. This can arise either from the geometry of the lattice under
consideration or from the introduction into the Hamiltonian of competing
exchange interactions. An example of the former is the basic Heisenberg
model on a 2D triangular lattice; while an example of the latter arises 
from the introduction on a bipartite lattice of (antiferromagnetic) 
next-nearest-neighbour interactions in addition to the (antiferromagnetic)
nearest-neighbour interactions of the pure Heisenberg model, resulting in 
the so-called $J_1$--$J_2$ model studied here.

Relatively few QMC calculations on such frustrated spin-lattice systems have
been performed. As a starting point they typically require a good trial wave
function, in terms of which the true wave function is well approximated,
especially for its nodal surface structure. In such calculations there can still be
a considerable systematic uncertainty, beyond the unavoidable statistical errors,
which arises from whether the simulations have eliminated the bias inherent in
the starting function. A typical recent calculation of this type was the fixed-node
Green function Monte Carlo method [6] simulation of the spin-half 
2D triangular-lattice
Heisenberg antiferromagnet by Boninsegni [7]. \ While undoubtedly representing
a very ambitious calculation of its kind, the resulting prediction for the sublattice
magnetisation, which is the simplest measure of the antiferromagnetic N\'{e}el
long-range order in this system, seems to be clearly too high by comparison with
the results from the best of the alternative techniques, including exact 
diagonalisations of small clusters [2] and the coupled cluster method [8]. \
Furthermore, even the resulting QMC estimate for the ground-state energy of the
triangular Heisenberg antiferromagnet gives an upper bound which is relatively
easily bettered by the alternative techniques.

For such frustrated and similarly ``difficult'' systems, predictions based on 
even very large-scale QMC simulations share, at least to some extent, the 
uncertainties discussed above for variational calculations. In order to overcome
these uncertainties, therefore, there is a real need to apply alternative
semi-analytical approaches, especially those that have the demonstrated power
to provide accurate predictions for the quantum order and for the positions and
critical properties of any quantum phase transitions. One such method, namely
the coupled cluster method (CCM) [9-19], stands to the fore in this respect. It
has long been acknowledged as providing one of the most powerful, most
widely applicable, and numerically most accurate at attainable levels of
computational implementation, of all available {\it ab initio} formulations
of microscopic quantum many-body theory. Furthermore, in recent years it has
been widely applied to many spin-lattice Hamiltonians [8,20-33]. \ For 
example, very successful applications have by now been made to the solid
phases of $^3He$ [20]; the isotropic Heisenberg and anisotropic $XXZ$
models in 1D and on the 2D square lattice,  both for spin-half systems [20-28]
and higher-spin systems [29]; and the spin-one Heisenberg-biquadratic 
model on the 1D chain [30]; as well as to such frustrated spin-half models as
the $J_1$--$J_2$ model in 1D (and 2D) [31-33] and the 2D triangular-lattice
anisotropic Heisenberg antiferromagnet [8,26,28].

In the present paper we apply the CCM specifically to investigate phase
transitions in the spin-half $J_1$--$J_2$ model on 
(the 1D chain and, especially,
on) the 2D square lattice. Our main aim is to use this model as an archetypal
example for which no exact information is known for the nodal structure of
the exact ground-state wave function, apart from the Marshall-Peierls 
sign-rule results in the pure Heisenberg limit. There is particular interest in
studying whether the sign rule is preserved when weak next-nearest-neighbour
exchange interactions are included and, if so, whether there is a critical
coupling beyond which the sign rule breaks down. We believe that the CCM is
an excellent {\it prima facie} candidate for such studies, 
since in virtually all
previous applications to models for which the Marshall-Peierls sign rule holds,
the theorem is exactly obeyed at virtually all levels of implementation in 
different CCM approximation schemes.

Finally, we are also interested in examining the phase transition points, as the
strength of the next-nearest-neighbour interactions is varied, at which the 
N\'{e}el antiferromagnetic long-range order present in the 2D square-lattice
case at the Heisenberg point vanishes; and in studying whether there is any
relationship between the phase boundaries and the points at which the 
Marshall-Peierls sign rule breaks down. We note that any reliable information
gained on the pattern of the signs of the multi-spin cluster coefficients in the
decomposition of the ground-state wave function should be very useful for two
distinct reasons, namely (i) for use in devising improved trial starting wave
functions for future QMC calculations, and (ii) for spotting possible patterns
for the signs of the cluster coefficients in different phases or different regimes
of coupling constants. The latter could be used, {\it inter alia,} to suggest
possible generalisations of the Marshall-Peierls sign rule, and thereby to 
motivate the search for the proofs of suitably generalised theorems. Any such
generalisations would clearly have immediate impact for a next generation of
QMC calculations.

The rest of this paper is organised as follows. In Sec.\ II we discuss the 
Marshall-Peierls sign rule and the CCM formalism. The sign rule is first
outlined for the Heisenberg model on bipartite lattices, before we describe the
$J_1$--$J_2$ model and previous results. The CCM formalism is then reviewed
in very general terms before describing one means of applying it to the 
$J_1$--$J_2$ model. Results for the ground-state energy and staggered 
magnetisation are discussed in Sec.\ III, together with results on the 
breakdown of the Marshall-Peierls sign rule for the model. Finally, we
present our conclusions in Sec.\ IV.


\section{The Marshall-Peierls Sign Rule and the CCM formalism}

\subsection{The Sign Rule For The Spin-Half
Heisenberg Antiferromagnet}

In this section we consider the spin-half Heisenberg 
antiferromagnet (HAF) on a bipartite lattice, where the Hamiltonian
is given by,
\begin{equation}
H=\sum_{\langle i,j \rangle}^N {\bf s}_i . {\bf s}_j ~ ,
\label{eq1} 
\end{equation}
and the index $i$ runs over all lattice points and $j$ runs
over the nearest neighbours to $i$. The angular brackets indicate
that we count each nearest-neighbour bond only once. We note that for
a bipartite lattice we can divide the lattice into two 
sublattices such that if $i$ is on one particular sublattice 
then $j$ must be on the other and vice versa.

For the one-dimensional (1D) linear chain, there is an exact 
solution to this model via the Bethe Ansatz technique [1].
For the two-dimensional (2D) square-lattice HAF, there is 
no exact solution to this problem, though many approximate
calculations, including those using various quantum Monte 
Carlo [3,4] (QMC) methods and exact series expansion [34]
techniques, have been performed.

Although no exact solution is known for the 2D
case stated here we note that there is an exact 
sign rule first derived by Marshall [5]
(and which we shall refer to here as the Marshall-Peierls
sign rule). The rule for the square lattice HAF 
is in fact an illustration of the more general Marshall-Peierls
sign rule for the HAF on any bipartite
lattice. This sign rule provides exact information
regarding the signs of the expansion coefficients
of the ground-state wave function in an Ising
basis, which is denoted $\{|I\rangle \}$. The exact 
ground-state wave function for an $N$-body spin
system may be written as,
\begin{equation}
| \Psi \rangle = \sum_I \Psi_I |I\rangle ~ ,
\label{eq2} 
\end{equation} 
where $\{\Psi_I \}$ are the expansion coefficients. 
We now divide the bipartite lattice into
its two sublattices, denoted $A$ and $B$, such that each 
nearest-neighbour site to an $A$ sublattice site
is on the $B$ sublattice and vice versa. If the number
of up spins on the $A$ sublattice is called $p_I$ then 
it is possible to show [5] that 
the coefficients $\{\Psi_I \}$ satisfy
\begin{equation}
\Psi_I = (-1)^{p_I} a_I,
\label{eq3}
\end{equation}
where the new coefficients $a_I$ are all positive. This 
exact information regarding the signs of the
coefficients may be used to
define the nodal surface of the wave function in this
basis, and hence is of use in QMC calculations [3,4].

\subsection{The $J_1$--$J_2$ Model}

We shall now discuss the  spin-half $J_1$--$J_2$ 
model on the 1D linear chain and the 2D square lattice.
The Hamiltonian is given by,
\begin{equation}
H=J_1 \sum_{\langle i,j \rangle}^N {\bf s}_i . {\bf s}_j 
+ J_2 \sum_{\langle \langle i,k \rangle \rangle}^N {\bf s}_i . {\bf s}_k ~ ,
\label{eq4} 
\end{equation}
where the sum on $\langle i,j \rangle$ runs over all 
nearest-neighbour pairs of sites, counting each pair (or bond)
once and once only; and the sum on $\langle \langle i,k \rangle 
\rangle$ similarly runs over all next-nearest-neighbour 
pairs of sites, again counting each pair (or bond) once and once only. 
We note that in order to consider a wide range of the coupling
parameters $J_1$ and $J_2$, it is useful 
to introduce the variable $\omega$ such that 
$J_1 \equiv {\rm cos} ~ \omega$ and 
$J_2 \equiv {\rm sin} ~ \omega$.

In 1D, no exact solution has been found for general values
of the coupling constants $J_1$ and $J_2$, though there
are some exact solutions including the Heisenberg point
($J_2=0$) and a point at $J_2/J_1=0.5$ at which
the ground-state is fully dimerised [35]. Previous
coupled cluster method (CCM) [31-33]  
and density matrix renormalisation group (DMRG) [9] 
calculations have very successfully been carried 
out for this model. The phase diagram 
is complicated, with  three distinct phases.
These phases may be characterised for our purposes as
ferromagnetic, antiferromagnetic, and frustrated.
The ferromagnetic phase is a highly degenerate 
phase in which the ground-state energy is 
equal to that of the 
classical fully aligned state. There is a first-order
phase transition at $J_1=0$ with negative $J_2$
to an antiferromagnetic phase. The antiferromagnetic
phase classically has its energy minimised by the
N\'eel state, and the quantum-mechanical phase
transition point to the frustrated phase is at 
(or is very near to) $J_2/J_1=0.5$. The 
frustrated phase classically contains a spin 
`spiral' state which has a periodicity which varies 
with the ratio of the coupling constants
$J_2/J_1$. There is some evidence that this 
changing periodicity with $J_2/J_1$ might also 
be seen in the quantum-mechanical 
system [33,36].

For the square lattice there are no
exact results, though approximate spin
wave theory (SWT) [37] calculations,
exact diagonalisations of finite-sized
lattices [38], and CCM [31]
calculations have been performed. The 
ferromagnetic to antiferromagnetic 
phase transition point is, as for the 1D
case, at $J_1=0$ with negative $J_2$,
and the antiferromagnetic to frustrated
phase transition is believed to be near to 
$J_2/J_1=0.5$.

The Marshall-Peierls sign rule, as discussed
in Sec. II.A, is true for the Heisenberg model
on a bipartite lattice. It is simple to 
prove that it is also preserved for the 
$J_1$--$J_2$ model with negative $J_2$ and positive
$J_1$. However, it is not in general 
true for positive $J_2$ and positive $J_1$. 
In fact, the results from  1D short-chain calculations [36]
suggest that the breakdown occurs very
near to the Heisenberg point, at 
$J_2/J_1 = 0.032 \pm 0.003$. By contrast,
finite-size lattice calculations [38] on the 
square lattice indicate that the sign 
rule at the Heisenberg point may well 
be preserved up to some critical value of $J_2/J_1$ 
in the $0.2 \leq J_2/J_1 \leq 0.3$.

\subsection{The CCM Formalism}

In this article we wish to perform CCM calculations
for the $J_1$--$J_2$ model in the antiferromagnetic 
regime. We now present a brief survey of the CCM 
formalism and note that a much fuller account of the formalism
as applied to quantum spin-lattice problems has
been  given in Ref. [8]. A more extensive overview of the
method and its applications has also been given 
in Ref. [17]. The starting point 
for any CCM calculation is the choice of a normalised model
or reference state, denoted $|\Phi\rangle$. We 
define a complete set of mutually commuting, 
multi-spin creation
operators $\{ C_I^{+} \}$ with respect to 
$|\Phi\rangle$ such that any Ising state $|I\rangle$ 
may be obtained as 
\begin{equation}
|I\rangle \equiv C_I^{+} |\Phi\rangle,
\label{eq5}
\end{equation}
for an $N$-body spin system. The ground-state wave function
has previously been written in Eq. (\ref{eq2}) as a 
linear combination of the states $\{ |I\rangle \}$, and we
now introduce the usual CCM parametrisations of the 
ket and bra ground states which are given by,
\begin{eqnarray}
|\Psi\rangle = e^S|\Phi\rangle &;& S=\sum_{I\neq 0}
{\cal S}_IC^+_I  ~~;~~~~  \label{eq6} \\
\langle\tilde{\Psi}| =  \langle\Phi |\tilde{S}e^{-S} &~~;~~~~&
\tilde{S}=1+\sum_{I\neq 0}\tilde{\cal S}_IC^-_I ~ .
\label{eq7}
\end{eqnarray}
The ket-state correlation operator in Eq. (\ref{eq6}) is,
as we can see, formed from a linear combination of the creation 
operators $\{ C_I^{+} \}$  multiplied with the 
relevant ket-state correlation coefficients $\{ S_I \}$.
The Hermitian adjoints of the multi-spin operators $\{ C_I^{+} \}$
are the multi-spin destruction operators $\{ C_I^- \}$, and the 
bra state in Eq. (\ref{eq7}) is formed by the linear combination of 
these destruction operators multiplied with the corresponding bra-state 
correlation coefficients $\{ \tilde S_I \}$. The 
bra and ket states, {\it defined} by Eqs. (\ref{eq6})
and (\ref{eq7}), are not manifestly Hermitian adjoints
of each other and so the variational property of an upper
bound on the ground-state energy is not preserved. However,
we note that the Hellmann-Feynman theorem is preserved. We 
also note that since $\langle \Phi| C^+_I = 0 = C^-_I | \Phi 
\rangle$ by definition, we have the explicit normalisation
relations, $\langle \Phi| \Psi \rangle = \langle \tilde \Psi | 
\Psi \rangle =   \langle \Phi| \Phi \rangle = 1$.

The ground-state expectation value of the energy 
may now simply be written using the Schr\"odinger
equation, $H|\Psi\rangle=E_g|\Psi\rangle$, as,
\begin{equation}
E_g = \langle \Phi | e^{-S} H e^S | \Phi \rangle ~ .
\label{eq8}
\end{equation}
Equation (\ref{eq8}) shows an example of 
the well-known similarity transform which plays
a crucial role in the CCM formalism. We further
note that the similarity transform of any 
quantum mechanical operator may be written 
in terms of a series of nested commutators, 
so that for the Hamiltonian $H$ we have
\begin{equation}
e^{-S} H e^S = H + [H,S] + \frac 1{2!} [[H,S],S] + 
\cdot\cdot\cdot ~~~.
\label{eq9}
\end{equation}
The infinite series of Eq. (\ref{eq9}) 
terminates at finite order if the Hamiltonian $H$ 
contains sums of products of only 
finite numbers of single-body operators, as 
is almost always the case and is, indeed, true for
the model considered here. We also
note that each time we perform a commutation operation in Eq. 
(\ref{eq9}) we produce a link or contraction, so that 
every single operator in each $S$ within 
the nested commutator expansion is directly linked to an operator
in the original Hamiltonian. In this way the Goldstone
linked cluster theorem is satisfied 
and the expectation value of the energy, as well as all 
other expectation values, are size-extensive (i.e., they
are well defined in the asymptotic thermodynamic 
limit $N \rightarrow \infty$ at all levels of approximation
for the operator $S$). Indeed, the CCM
works from the outset in the thermodynamic limit.

We now wish to find values for the ket-state and 
bra-state correlation coefficients. We do this 
by defining the expectation value, $\bar H \equiv \langle 
\tilde \Psi | H | \Psi \rangle$, and by requiring that 
this quantity is a minimum with respect to the 
ket-state and bra-state correlation coefficients.
Hence, we have, 
\begin{eqnarray}
\delta\bar{H}/\delta\tilde{{\cal S}}_I=0\Rightarrow\langle\Phi |C^-_Ie^{-S}
He^S|\Phi\rangle =0~~&,&~~~~\forall I\neq 0~~; \label{eq10} \\
\delta\bar{H}/\delta{\cal S}_I=0\Rightarrow\langle\Phi |\tilde{S}e^{-S}
[H,C^+_I]e^S|\Phi\rangle =0~~&,&~~~~\forall I\neq 0~~. \label{eq11}
\end{eqnarray}

This formalism is exact in the limit that we include 
all possible multi-spin cluster correlations within $S$ and $\tilde S$,
though in any real application this is usually 
impossible. We therefore need to consider approximation schemes whereby the 
expansions of $S$ and $\tilde{S}$ in Eqs.\ (\ref{eq6}) 
and (\ref{eq7}) may be truncated to 
some finite or infinite subset of the full set of independent 
(fundamental) multi-spin configurations. The three
most commonly employed schemes have been: (1) the SUB$n$ scheme, in 
which all correlations involving only $n$ or fewer spins are retained, but no
further restriction is made concerning their spatial separation on the lattice;
(2) the SUB$n$-$m$  sub-approximation, in which all SUB$n$ 
correlations spanning a range of no more than $m$ adjacent lattice sites
are retained; and (3) the localised LSUB$m$ scheme, which retains all
multi-spin correlations over distinct locales on the lattice
defined by $m$ or fewer contiguous sites. 

In the next subsection we consider the application of the CCM 
to the $J_1$--$J_2$ model in the antiferromagnetic regime.

\subsection{The CCM Applied to the $J_1$--$J_2$ Model}

As stated in the previous section, the starting point
for any CCM calculation is the choice of the model (or reference)
state. Here, we choose the classical N\'eel state to be our model
state, in accordance with previous CCM calculations [8,31],
in order to study the antiferromagnetic regime of the $J_1$--$J_2$
model. We visualise the N\'eel state by again dividing 
the lattice into two sublattices, $A$ and $B$, on which 
each of the nearest-neighbours sites to a given 
sublattice site are on the other sublattice. We populate the 
$A$ sublattice with `up' spins (i.e., eigenvectors of the 
$s^z$ operator with eigenvalue $+\frac12$) and the $B$
sublattice with `down' spins (i.e., eigenvectors of the 
$s^z$ operator with eigenvalue $-\frac12$). 

In order to perform a CCM calculation we would like 
to treat each site on an equal footing.
We do this by performing a rotation [8,31,33] 
of the local axes of the spins 
on the $A$ sublattice (`up' spins) by $180^{\circ}$ 
about the $y$-axis such that all spins on each sublattice 
appear mathematically to point downwards (i.e., 
in these new local axes). Since this rotational 
transformation is a canonical one, it has no effect 
on the commutation relations. It does however have
a number of consequences. Firstly, the
Hamiltonian is re-written in the local coordinates as,
 \begin{eqnarray}
H &=& -J_1 \sum_{\langle i,j \rangle} \biggl [ 
s_i^z s_{j}^z +
\frac 12 s_i^+ s_{j}^+ +
\frac 12 s_i^- s_{j}^- 
\biggr ] \nonumber \\
& & + J_2 \sum_{\langle \langle i,k \rangle \rangle } \biggl [ 
s_i^z s_{k}^z +
\frac 12 s_i^- s_{k}^+ +
\frac 12 s_i^+ s_{k}^- 
\biggr ] \label{eq12} ~ .
\end{eqnarray}
We also note that the set of creation operators $\{ C_I^{+} \}$
may now be formed purely from products of spin raising operators
with respect to the rotated, `ferromagnetic' model state.
We write this expression for an $l$-spin cluster as 
$C_I^{+} \equiv s_{i_1}^+ s_{i_2}^+ ...s_{i_l}^+$.
Conversely, the destruction operators are now formed purely 
from the spin lowering operators in an analogous manner,
where $C_I^- \equiv s_{i_1}^- s_{i_2}^- ...s_{i_l}^-$. 

The Marshall-Peierls sign rule for the Heisenberg model
is also modified. We obtain a new and exact rule for the 
Hamiltonian of Eq. (\ref{eq12}) in an expansion of the ground-state
wave function in terms of an Ising basis, $\{ |I\rangle \}$,
in the local, rotated spin coordinates.  The corresponding 
expansion coefficients, $\{ \Psi_I \}$, must now be positive for 
all of the states labelled by $I$. (A proof of this statement is not
given here, but it is made in 
exactly the same manner as that of Marshall [5].)
The $\{ \Psi_I \}$ coefficients are henceforth
explicitly stated in relation to the Ising basis in the
local, rotated spin coordinates.

We now wish to provide a link between the $\{ \Psi_I \}$
coefficients, in terms of the local axes, and the CCM
ground-state parametrisation of the ket state of Eq. 
(\ref{eq6}).  This is done by applying the destruction
operator $C_I^-$, for a particular cluster defined by the index $I$,
to the expressions for the ket-state
wave function of Eqs. (\ref{eq2}) and (\ref{eq6}). 
Note we choose only one ordering out of the indices
$\{i_1,i_2,...,i_l\}$ of the total of $N(l!)\nu$ 
possible equivalent orderings for $C_I^-$ on the lattice,
where $\nu$ is a symmetry factor dependent on the lattice.
We therefore write the $\{ \Psi_I \}$ coefficients as, 
\begin{equation}
\Psi_I = \langle \Phi | C_I^- e^S | \Phi \rangle ~ \equiv ~
\langle \Phi | s_{i_1}^- s_{i_2}^- ...s_{i_l}^- ~
e^S | \Phi \rangle ~ .
\label{eq13}
\end{equation}
Note that Eq. (\ref{eq13}) contains the implicit assumption
that the spin raising operators in $C_I^{+}$
of Eq. (\ref{eq5}), which are used to define $|I\rangle$ with 
respect to $|\Phi\rangle$, have only one ordering with respect to 
permutations of the indices $\{i_1,i_2,...,i_l\}$.

Again, it should be noted that in
practice one restricts the choice of the 
clusters contained within $S$ to some well-defined 
approximation scheme. To keep the calculations
as self-consistent as possible, we restrict the 
choice of the $\{ \Psi_I \}$ coefficients to be
for only those Ising states defined in Eq. (\ref{eq5})
which correspond to the clusters used in $S$. 

In the next section we describe our results for 
the ground-state expectation values for high-order, 
approximate CCM calculations which are determined 
computationally [8]. We also detect 
critical points in the CCM equations which
are taken to be signatures of phases transitions
in the real system. Once the ket-state 
correlation coefficients are found it is then 
possible to obtain approximate results for the  
$\{ \Psi_I \}$ coefficients, again via a 
computational approach, and we discuss 
CCM results concerning the breakdown of
the Marshall-Peierls sign rule as a function
of $J_2/J_1$.


\section{Results}

\subsection{Ground-State Expectation Values}


The ground-state energy of Eq. (\ref{eq8}) is approximately
obtained once the CCM equations are first derived and then 
solved for a particular approximation scheme and approximation
level. Descriptions of the method are 
given in Refs. [31-33]. Details of how 
one may obtain a computational solution
for high-order LSUB$m$ approximations is given in 
Ref. [8]\footnote{It should be noted that
the calculation of Ref. [31] was mostly 
concerned with SUB2 calculations for the 
spin-half $J_1$--$J_2$ model. However, a
calculation for the square lattice system
in which only nearest-neighbour correlations and 
four-body correlations between four spins on the unit square 
were retained was also performed. This calculation was 
referred to as an `LSUB4' calculation within 
the text in this reference to denote the addition of
the extra, single type of four-body correlation.
However, this `LSUB4' calculation was not
the same as the LSUB4 calculation which we perform
here which now contains {\it all} two-body and 
four-body correlations in a locale defined 
by $m$=4.}. We simply quote the results here
for this model using the N\'eel model state and the
interested reader is referred to these articles.

\begin{figure}
\epsfxsize=8cm
\centerline{\epsffile{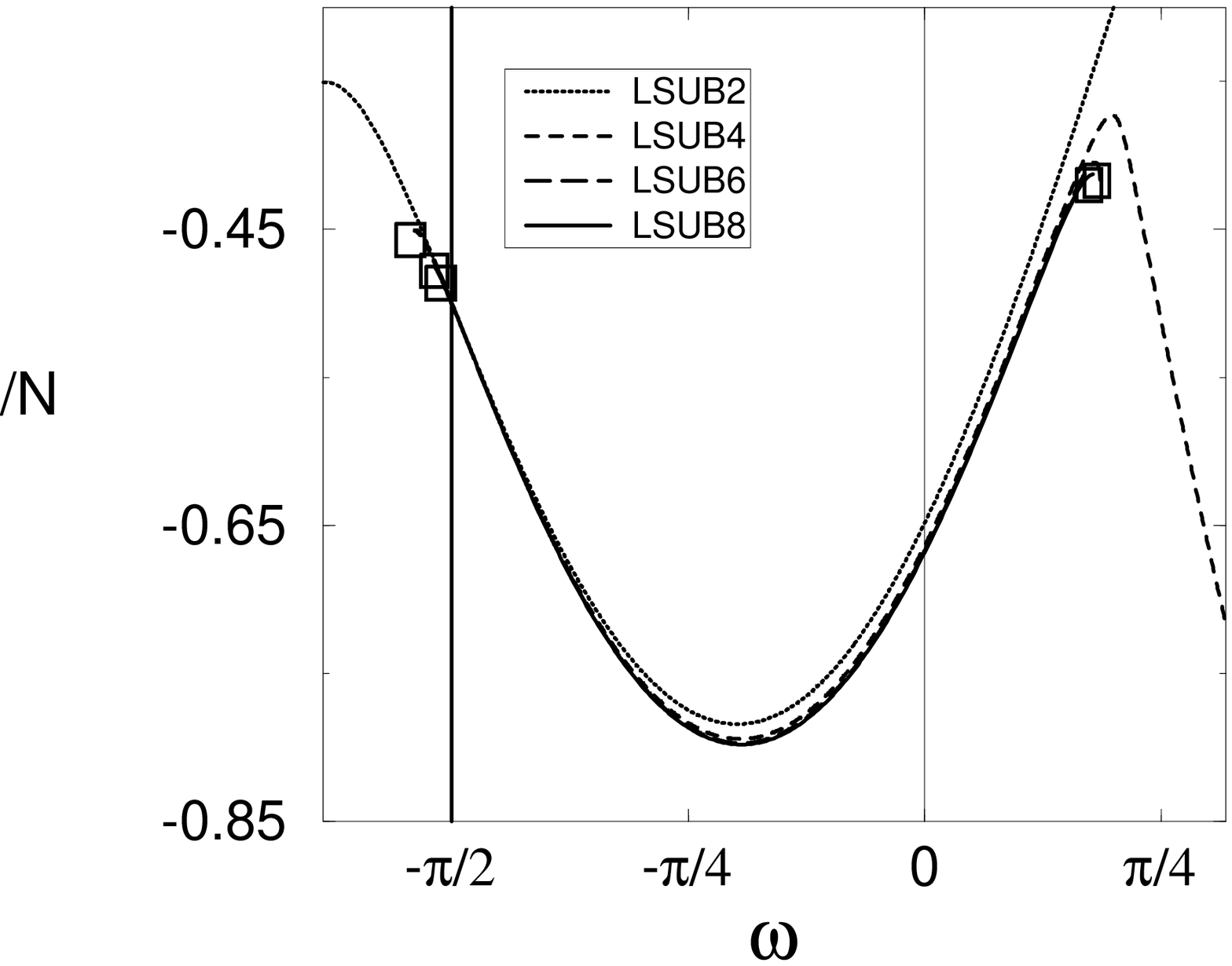}}
\caption{Results for the CCM ground-state energy of the spin-half
$J_1$--$J_2$ model on the 2D square lattice, using the LSUB$m$ approximation
based on the N\'{e}el model state, with $m=2,4,6,8$. LSUB$m$ critical points,
$\omega_F$ and $\omega_A$, are indicated by the boxes. Note that 
$J_1\equiv {\rm cos} ~ \omega $ and $J_2 \equiv {\rm sin} ~ \omega$.}
\label{fig1}
\end{figure}

The LSUB$m$ results for the ground-state energy of the 1D
$J_1$--$J_2$ model converge very well over the
range $-\pi/2 \leq \omega \leq {\rm tan}^{-1} (0.5)$.  
We note that the LSUB10 results agree to within 1$\%$ of 
those obtained by extrapolating the results 
from exact diagonalisations for short chains
[31] over this range,
though we do not provide a plot of this here.
In 2D, we see in Fig. \ref{fig1} that our results are 
again extremely well converged over the range $-\pi/2 \leq 
\omega \leq {\rm tan}^{-1} (0.5)$. 
In Table \ref{tab1} results are given for the 
ground-state energy of the square lattice
system as a function of $\frac {J_2}{J_1} \equiv 
{\rm tan} \omega$ for $-0.5 \leq \frac {J_2}{J_1} 
\leq 0.5$ for the LSUB6 and LSUB8 levels 
of approximation.


We note that in 2D the CCM results for the ground-state energy 
display characteristic terminating points at certain
critical values of $\omega$. At these points 
the second derivative of the ground-state energy
with respect to $\omega$ may also be determined,
and we note that at these {\it critical} values
of $\omega$ this quantity diverges. This type of 
behaviour has been observed previously [8]
and is associated with a phase 
transition in the real system. 
The critical value in 2D near to
the ferromagnetic phase transition, denoted
$\omega_F$, is given in Table \ref{tab2}. 
We see that the LSUB$m$
results are clearly converging to the exact value
of $\omega_F=-\pi/2$. It is known [31] that 
the SUB2 approximation predicts this point exactly
in both 1D and 2D. 

As is seen from the entries in Table II, the 
antiferromagnetic point, denoted $\omega_A$, 
is detected in 2D with the LSUB6, LSUB8, and SUB2 approximations.
It is not observed within the LSUB4 approximation.
We can see that the LSUB$m$ critical value of $\omega$
decreases with increasing truncation index $m$, and an 
simple extrapolation [8] in the limit $m \rightarrow \infty$
gives a value for the phase transition point of 
$ \frac {J_2}{J_1} \approx 0.61$.


We now introduce the sublattice magnetisation, which characterises 
the degree of quantum order inherent in the CCM wave functions.
By inserting the CCM parametrisations of Eqs. (\ref{eq6}) 
and (\ref{eq7}) we find,
\begin{equation}
M\equiv-\frac{2}{N}\sum^N_{k=1}\langle\tilde{\Psi}|s^z_k|\Psi
\rangle =-\frac{2}{N}\sum^N_{k=1}\langle\Phi |\tilde{S}
e^{-S}s^z_ke^S|\Phi\rangle ~~,
\label{eq14}
\end{equation}
where $s^z_k$ is in the local coordinates of each sublattice. 
Evaluation of the sublattice magnetisation requires both the ket- and 
bra-state cluster correlation coefficients. The actual procedure to do 
this is straightforward, and is also described in more
detail elsewhere [8].

The sublattice magnetisation in 1D is non-zero in the range
$-\pi/2 \leq \omega \leq {\rm tan}^{-1} (0.5)$, though we
note (see [31]) that it is greater than
zero but monotonically decreases with increasing LSUB$m$ 
approximation level for all $\omega$ in this range. The sublattice
magnetisation is zero in the true solution of this 
model in 1D, and although the CCM LSUB$m$ (and SUB2 [31]) 
results are non-zero we expect that with increasing 
level of LSUB$m$ approximation this would be better reflected in the 
CCM solution.

\begin{figure}
\epsfxsize=8cm
\centerline{\epsffile{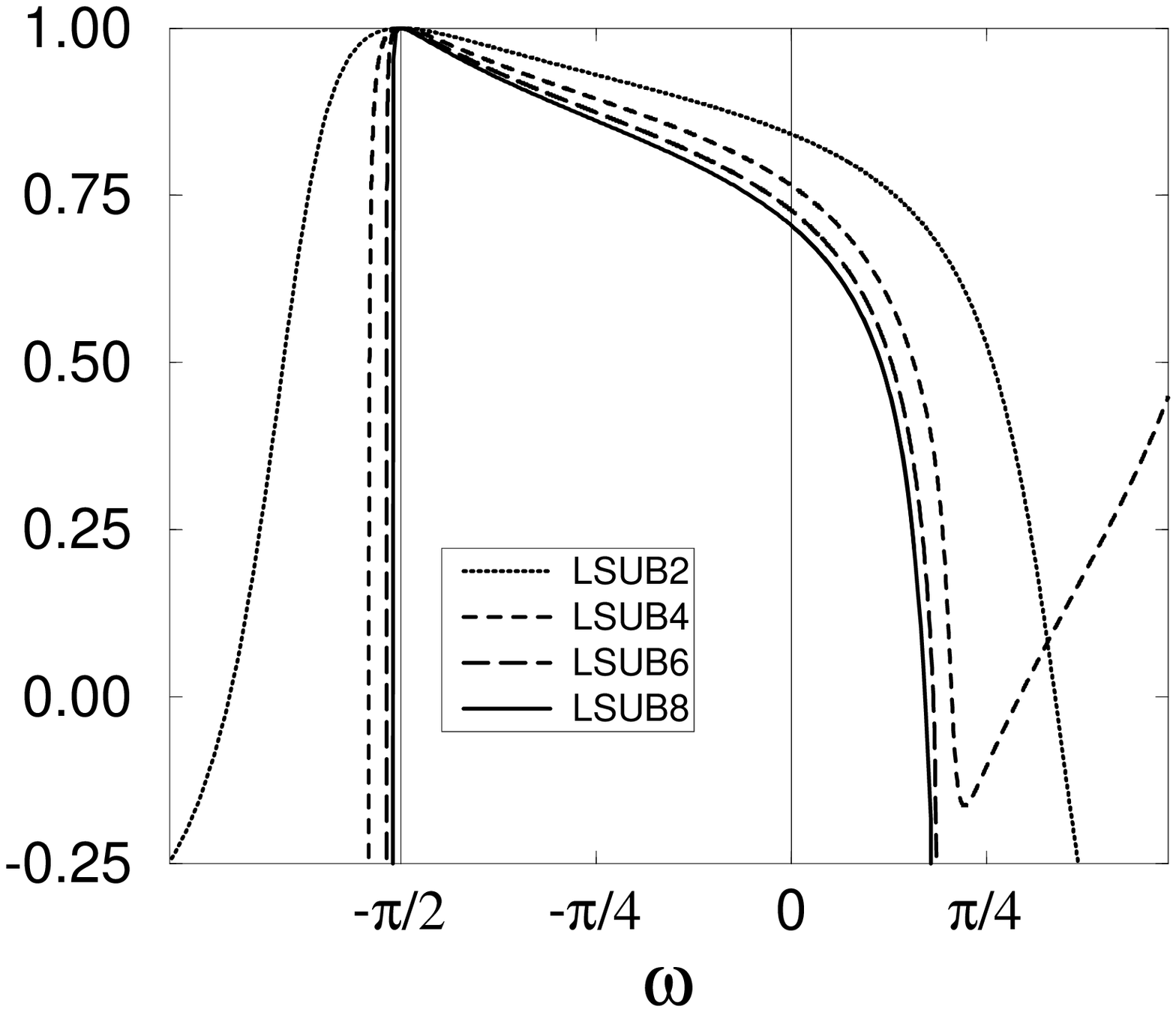}}
\caption{Results for the CCM sublattice magnetisation of the spin-half
$J_1$--$J_2$ model on the 2D square lattice, using the LSUB$m$ approximation
 based on the N\'{e}el model state, with $m=2,4,6,8$. Note that 
$J_1 \equiv {\rm cos} ~ \omega$ and $J_2 \equiv {\rm sin} ~ \omega$.}
\label{fig2}
\end{figure}

Figure \ref{fig2} illustrates that that the situation 
is much clearer in 2D. We can see that the sublattice 
magnetisation in Fig. \ref{fig2} is converging to a 
non-zero value over essentially all of the 
range $-\pi/2 \leq \omega \leq {\rm tan}^{-1} (0.5)$. 
We note that there are divergences in the sublattice 
magnetisation which are observed at precisely the same points 
as the critical points, $\omega_F$ and $\omega_A$, 
of the energy in 2D. This reinforces our conjecture 
that these critical points are reflections
of phase transitions in the real system.

\subsection{The Breakdown of the Marshall-Peierls Sign Rule}

We now consider the Marshall-Peierls sign 
rule for the spin-half $J_1$--$J_2$ model
on the linear chain and the square lattice. 
We need to obtain the $\{ \Psi_I \}$ coefficients, 
either analytically or computationally, 
in terms of the ket-state coefficients. 
We then solve the SUB2 or LSUB$m$ equations in order 
to obtain the ket-state correlation coefficients
and hence to obtain approximate values for the 
$\{ \Psi_I \}$ coefficients. Note that these
calculations are approximate in the sense that we
only retain certain correlations in $S$ with a 
well-defined approximation scheme, though we 
are {\it already} working in the infinite 
lattice or $N \rightarrow \infty$ limit.

We note that at each order of LSUB$m$ approximation it
is possible to perform this process of matching the terms 
in $e^S$ in Eq. (\ref{eq13}) to the configuration $C_I^-$ 
analytically. In Appendix A we present the exact form of the 
$\{ \Psi_I \}$ coefficients in terms of the ket-state 
coefficients within the LSUB4 approximation scheme 
for the 1D linear chain. Furthermore, we see that each 
of the $\{ \Psi_I \}$ coefficients corresponding
to two-body correlations with respect to $|\Phi\rangle$
via Eq. (\ref{eq5}), is exactly equal to the corresponding
CCM two-body ket-state correlation operator $\{ S_I \}$
for any approximation scheme used in $S$. This may be 
proven by noting that there are no one-body spin correlations 
allowed in $S$ (in order to preserve the conserved quantity 
$s_T^z = \sum_i^N s_i^z ~ $), and then by considering the
series expansion of the exponential in Eq. (\ref{eq13}).

To obtain higher-order $\{ \Psi_I \}$ coefficients
using Eq. (\ref{eq13}) we may conveniently use a computational 
approach. This amounts to partitioning the configuration 
in $C_I^- \equiv s_{i_1}^- s_{i_2}^- ... s_{i_l}^-$ into 
the multiples of $S$ in the series expansion of $e^S$ of 
Eq. (\ref{eq13}). It is then possible to identify by simple
computer algebra 
the configurations of the partitioned pieces (each referring 
to an $S$ in the series expansion of the exponential), 
and find a numerical value for the $\{ \Psi_I \}$ 
coefficients once the CCM ket-state equations have 
been solved at specific values of $\omega$.

In both 1D and 2D, we find that (at all levels of approximation)
the Marshall-Peierls sign rule is preserved for 
$\omega=-\pi/2|_+$ (i.e., in the antiferromagnetic
regime). That is, all of the $\{ \Psi_I \}$ 
coefficients are found to be positive. The sign is broken 
for $\omega=-\pi/2|_-$ (i.e., in the ferromagnetic regime)
at which point at least one of the coefficients becomes 
negative. Note that the crossover occurs exactly at 
the phase boundary $\omega=-\pi/2$, and that this is
one case where the breakdown of the Marshall-Peierls
sign rule occurs at exactly the same place 
as the phase boundary. We note that there is
a first-order phase transition at this point.

In 1D, we find that all the $\{ \Psi_I \}$ coefficients
are positive at the Heisenberg point for the SUB2 scheme
and for LSUB$m$ schemes with $m \leq 8$. 
Above this LSUB$m$ level of approximation (i.e., for $m > 8$)
a few of the $\{ \Psi_I \}$ coefficients (which are  
very small in magnitude) become negative.
However, for example, at the LSUB12 level we find that 
these same coefficients, which are negative at the LSUB10 level,
again become positive, though we 
find that other new coefficients (which are similarly 
small in magnitude) then become negative. 
These would in turn presumably become 
positive at a still higher level of approximation. Thus, 
the CCM is completely consistent with the sign rule
for the Heisenberg model in 1D.
This is encouraging as this model is quite 
challenging for the CCM with this model state, 
as our results for the sublattice magnetisation
have shown. We may also compare the ratios of the magnitudes 
of the $\{ \Psi_I \}$ coefficients to those obtained
via short-chain calculations, as shown in Table \ref{tab3}. 
(Note that we examine the ratios to eliminate 
short-chain normalisation 
factors.) We can see that the correspondence
between CCM and short-chain calculations 
is good, though it appears that the CCM
results are better converged at the LSUB10
and LSUB12 levels of approximation than those from the
12-spin and 16-spin chains. Short-chain
calculations [36] indicate 
the breakdown of the Marshall-Peierls
sign rule at $J_2/J_1 = 0.032 \pm 0.003$,
though our CCM results cannot give an 
accurate value for this breakdown point.

For the square lattice, the situation is 
found to be much clearer.
The signs of the $\{ \Psi_I \}$ coefficients
are found to be positive at the Heisenberg
point at all orders of LSUB$m$ approximation 
and also from the SUB2 approximation. 
A clear transition from all of 
the coefficients being positive to one of them 
becoming negative is seen, and we believe
that this clearly indicates the onset of the breakdown 
of the Marshall-Peierls sign rule. The points 
at which the LSUB$m$ approximation predicts a 
breakdown, denoted $\omega_M$, 
are shown in Table \ref{tab2}.
We can see that a simple extrapolation 
of these points gives a value for
the breakdown of the sign rule to be
at $J_2/J_1 \approx 0.26$
which is in good agreement with the 
SUB2 result of $J_2/J_1 = 0.2607$
and exact diagonalisations of finite-sized 
lattice calculations [38]
which give a corresponding value of $J_2/J_1$
in the range of $0.2$--$0.3$.
At SUB2 and LSUB4 levels of 
approximation it is the $\{ \Psi_I$ \}
coefficients for the two-spin cluster 
with separation $\hat x + 2 \hat y$ that 
becomes negative first, where $\hat x$ and 
$\hat y$ are the unit vectors along the perpendicular 
axes of the square lattice. Higher-order CCM 
LSUB$m$ calculations predict, however, that the 
coefficients for other higher-order, highly-disconnected 
configurations first become negative at a slightly 
lower value of $\omega$ than the coefficient for
this two-body configuration. This therefore
indicates that higher-order multi-spin configurations
might as be important as this two-body
configuration in the breakdown of the 
sign rule for this model.

We note that the square-lattice results 
predict that the breakdown of the sign rule 
($\omega_M$) occurs at a smaller value of $\omega$,
at a particular approximation level, than 
the antiferromagnetic critical point 
($\omega_A$) predicted by the CCM. 
In other words, the CCM results predict that there 
is a region of the antiferromagnetic regime
in which the Marshall-Peierls sign rule 
is not being obeyed.


\section{Conclusions}

The CCM applied to the 1D spin-half $J_1$--$J_2$ model gives encouraging 
results for the ground-state energy, the critical points and the sublattice
magnetisation. In this paper we have shown how the recent advances in the 
computational implementation of the method enable us to obtain useful 
results for the 2D model as well. We find that the 2D case is in many 
ways simpler than the 1D case, with more clearly defined critical points
and a non-zero sublattice magnetisation.

We have also investigated the relation between the CCM and the 
Marshall-Peierls sign rule. In 1D we find the CCM results are consistent 
with the exact results in the region where these apply. In 2D we are able 
to obtain results which are better than the finite-size extrapolations
and we can predict the point at which the sign rule fails. Our results 
indicate that this occurs at a different point than the phase transition 
from the simple antiferromagnet to the more complicated frustrated phase.
We believe that the use of the exact Marshall-Peierls sign rules, 
extended by the CCM method into regions where it is not exact, can 
shed new light on the behaviour of this type of quantum spin system.
Furthermore, it may provide information about the nodal surface that 
can be used for accurate QMC calculations in a much wider range of 
quantum spin systems than previously.

\section*{Acknowledgements}

One of us (RFB) also gratefully acknowledges a research grant from the
Engineering and Physical Sciences Research Council (EPSRC) of Great
Britain. This work has also been supported in part by the Deutsche
Forschungsgemeinschaft (GRK 14, Graduiertenkolleg on `Classification of 
phase transitions in crystalline materials').

\pagebreak

\appendix

\section{Analytic Evaluation of the $\Psi_I$ Coefficients}

We now present analytical expressions for the $\{ \Psi_I \}$ coefficients
in terms of the ket-state correlation coefficients in $S$ 
for the LSUB4 approximation in 1D. The ket-state correlation 
operator for the LSUB4 approximation in 1D is given by,
\begin{equation}
S = b_1 \sum_i^N s_i^+ s_{i+1}^+ + b_3 \sum_i^N s_i^+ s_{i+3}^+ + 
g_4 \sum_i^N s_i^+ s_{i+1}^+ s_{i+2}^+ s_{i+3}^+  
\label{a1} ~ ,
\end{equation}
where the index $i$ runs over all points on the linear chain. 
To obtain the $\{ \Psi_I \}$ coefficients we now must 
choose the $C_I^-$ configurations in Eq. (\ref{eq13}):
for the nearest-neighbour, two-body coefficient, which we 
shall denote as $\Psi_1$, we use $C_1^- = s_j^- s_{j+1}^-$; 
for the third-nearest-neighbour, two-body 
coefficient, which we shall denote as $\Psi_2$,
we use  $C_2^- = s_j^- s_{j+3}^-$; and for the four-contiguous spin 
configuration coefficient, which we shall denote $\Psi_3$, we use 
$C_3^- = s_j^- s_{j+1}^- s_{j+2}^- s_{j+3}^-$. The result is
therefore,
\begin{eqnarray}
\Psi_1 & = & b_1 ~; \\
\Psi_2 & = & b_3 ~; \\
\Psi_3 & = & b_1^2 + b_1 b_3 + g_4 ~. 
\label{b2} 
\end{eqnarray}
The values of the $\Psi_I$ coefficients are independent of $j$, due 
to the translational symmetry of the lattice, and so
the index $j$ is chosen arbitrarily from any of its 
$N$ possible values in order to obtain Eqs. (A2-A4). 
Higher-order LSUB$m$ approximations can be handled 
analogously by making use of computer-algebraic
techniques.

\pagebreak


\pagebreak 


\pagebreak


\begin{table}
\caption{Results for the CCM ground-state energy 
of the spin-half $J_1$--$J_2$ model on the 
2D square lattice as a function of $J_2/J_1={\rm tan} ~ \omega$, 
using the LSUB$m$ approximation based on the 
N\'{e}el model state, with $m=6,8$.}
\begin{tabular}{|l|c|c|} 
$J_2/J_1={\rm tan}(\omega)$   &$E_g/N$~~($m$=6)  &$E_g/N$~~($m$=8)          
\\ \hline
$-$0.50               &$-$0.88237  &$-$0.88353    \\ \hline
$-$0.40               &$-$0.83785  &$-$0.83902    \\ \hline
$-$0.30               &$-$0.79393  &$-$0.79510    \\ \hline
$-$0.20               &$-$0.75071  &$-$0.75188    \\ \hline
$-$0.10               &$-$0.70833  &$-$0.70951    \\ \hline
0.00                  &$-$0.66700  &$-$0.66817    \\ \hline
0.10                  &$-$0.62699  &$-$0.62816    \\ \hline
0.20                  &$-$0.58868  &$-$0.58988    \\ \hline
0.30                  &$-$0.55271  &$-$0.55397    \\ \hline
0.40                  &$-$0.52012  &$-$0.52164    \\ \hline
0.50                  &$-$0.49311  &$-$0.49551        
\end{tabular}
\label{tab1}
\end{table}

\begin{table}
\caption{Results for the CCM critical points of the 
spin-half $J_1$--$J_2$ model on the 
2D square lattice. The ferromagnetic 
and antiferromagnetic critical points are denoted 
$\omega_F$ and $\omega_A$ respectively. The 
point ($\omega > 0$) at which the CCM predicts 
that the Marshall-Peierls 
sign rule breaks down for the square lattice 
is denoted $\omega_M$. Corresponding results
for $(J_2/J_1)|_A$ and $(J_2/J_1)|_M$ are also shown, 
where $J_2/J_1  ={\rm tan} ~ \omega$.}
\begin{tabular}{|c|c|c|c|c|c|}  
             &$\omega_F$         &$\omega_A$  &$(J_2/J_1)|_A$ 
&$\omega_M$  &$(J_2/J_1)|_M$      \\ \hline 
SUB2         &$-\pi/2$           &0.6416  &0.7470    &0.255  
&0.2607 \\ \hline
LSUB4        &$-$1.702           &--      &--        &0.331  
&0.344  \\ \hline
LSUB6        &$-$1.628           &0.583   &0.660     &0.290  
&0.298  \\ \hline
LSUB8        &$-$1.603            &0.566   &0.636     &0.275  
&0.282  \\ \hline
LSUB$\infty$ &$-$1.572            &0.544   &0.605     &0.255  
&0.261    \\ \hline
\end{tabular}
\label{tab2}
\end{table}

\begin{table}
\caption{Results for the ratios of the magnitudes of the
CCM LSUB$m$ $\Psi_I$ 
coefficients, for the spin-half, 1D Heisenberg model, compared to ratios 
of the magnitudes of the equivalent Ising expansion coefficients determined by 
finite-size, exact diagonalisations. The $\Psi_1$ coefficient 
is associated with the nearest-neighbour, two-body correlation 
with respect to the model state $| \Phi \rangle$; and $\Psi_2$, 
$\Psi_3$, and $\Psi_4$ are, respectively, the corresponding 
coefficients associated with the 3$^{\rm rd}$-nearest-neighbour, 
two-body configuration, the four-contiguous-spin configuration, 
and the six-contiguous-spin configuration.}
\begin{tabular}{|c|c|c|c|c|}  
 Ratio 
& 12 Spins  
& 16 Spins  
& LSUB10  
& LSUB12  \\ \hline
$\frac {|\Psi_3|} {|\Psi_1|}$ &0.7436       &0.7163
&0.6720 &0.6758 \\ \hline
$\frac {|\Psi_2|} {|\Psi_1|}$ &0.1674       &0.1552
&0.1413 &0.1381 \\ \hline
$\frac {|\Psi_4|} {|\Psi_1|}$ &0.4850       &0.6155
&0.5157 &0.5248 \\
\end{tabular}
\label{tab3}
\end{table}


\begin{thebibliography}{99}

\bibitem{ref1} H. A. Bethe, {\sl Z. Phys.} {\bf 71}, 205 (1931);
  L. Hulth{\'e}n, {\sl Ark. Mat. Astron. Fys. A}
  {\bf 26}, No. 11 (1938).

\bibitem{new1} B. Bernu, C. Lhuillier, and L. Pierre, 
{\sl Phys. Rev. Lett.} {\bf 69}, 2590 (1992); B. Bernu, 
P. Lecheminant, C. Lhuillier, and L. Pierre, 
{\sl Phys. Rev. B} {\bf 50}, 10048 (1994); P. Lecheminant, 
B. Bernu, C. Lhuillier, and L. Pierre, {\sl Phys. Rev. B} 
{\bf 52}, 9162 (1995).

\bibitem{ref2} J. Carlson,
  {\sl Phys. Rev. B} {\bf 40}, 846 (1989);
  N. Trivedi and D.M. Ceperley,
  {\it ibid.} {\bf 41}, 4552 (1990).

\bibitem{ref3}   K.J. Runge, 
  {\sl Phys. Rev. B} {\bf 45}, 12292 (1992);
  {\it ibid.} {\bf 45}, 7229 (1992). 

\bibitem{ref5} W. Marshall, {\sl Proc.\ R. Soc.\ London A}\ 
  {\bf 232}, 48 (1955).

\bibitem{new2} D.M. Ceperley and B.J. Alder, 
{\sl Phys. Rev. Lett.} {\bf 45}, 566 (1980);
Science {\bf 231}, 555 (1986); H.J.M. van 
Bemmel, D.F.B. ten Haaf, W. van Saarloos,
J.M.J. van Leeuwen, and G. An, 
{\sl Phys. Rev. Lett.} {\bf 72}, 2442 (1995).

\bibitem{new3} M. Boninsegni, 
                {\sl Phys. Rev. B} {\bf 52}, 15304 (1995). 

\bibitem{ref13} C. Zeng, D.J.J. Farnell, and R.F. Bishop, 
                {\sl J. Stat. Phys.}, {\bf 90}, 327 (1998).

\bibitem{new4} F. Coester,
                {\sl Nucl. Phys.} {\bf 7}, 421 (1958);  
                F. Coester and H. K\"ummel, {\it ibid.} {\bf 17}, 477 (1960).

\bibitem{new5} J. \v{C}i\v{z}ek,
                {\sl J. Chem. Phys.} {\bf 45}, 4256 (1966);   
                {\sl Adv. Chem. Phys.} {\bf 14}, 35 (1969).

\bibitem{new6} H. K\"ummel, K.H. L\"uhrmann, and J.G. Zabolitzky, 
                {\sl Phys Rep.} {\bf 36C}, 1 (1978).


\bibitem{new7} R.F. Bishop and K.H. L\"uhrmann,
                {\sl Phys. Rev. B} {\bf 17}, 3757 (1978);
                {\it ibid.} {\bf 26}, 5523 (1982). 

\bibitem{new8} J.S. Arponen, 
                {\sl Ann. Phys.} {\em (N.Y.)} {\bf 151}, 311 (1983).

\bibitem{new9} H.G. K\"ummel, in 
                {\sl Nucleon-Nucleon Interaction and Nuclear 
                Many-body Problems}, edited by S.S. Wu and
                T.T.S. Kuo (World Scientific, Singapore, 1984),
                p. 46.

\bibitem{new10} R.F. Bishop and H. K\"ummel,
                {\sl Phys. Today} {\bf 40(3)}, 52 (1987).

\bibitem{new11} J. Arponen, R.F. Bishop, and E. Pajanne, 
                {\sl Phys. Rev. A} {\bf 36}, 2519 (1987);
                {\it ibid.} {\bf 36}, 2539 (1987).

\bibitem{new12} R.F. Bishop, 
                {\sl Theor. Chim. Acta} {\bf 80}, 95 (1991). 

\bibitem{new13} R.F. Bishop, in {\sl Dirkfest '92 -- 
                A Symposium in Honor of J. Dirk Walecka's
                Sixtieth Birthday}, edited by W.W. Buck,
                K.M. Maung, and B.D. Serot (World
                Scientific, Singapore, 1992), p. 21.

\bibitem{new14} R.F. Bishop, in {\sl Many-Body Physics},
                edited by C. Fiolhais, M. Fiolhais,
                C. Sousa, and J.N. Urbano (World 
                Scientific, Singapore, 1994), p. 3.

\bibitem{new15} M. Roger and J.H. Hetherington, 
                {\sl Phys. Rev. B} {\bf 41}, 200 (1990).

\bibitem{new16} M. Roger and J.H. Hetherington, 
                 {\sl Europhys. Lett.} {\bf 11}, 255 (1990).

\bibitem{new17} R.F. Bishop, J.B. Parkinson, and Y. Xian, 
                  (a) {\sl Phys. Rev. B} {\bf 43}, 13782 (1991);
                  (b) {\sl Theor. Chim. Acta} {\bf 80}, 181 (1991);
                  (c) {\sl Phys. Rev. B} {\bf 44}, 9425 (1991);
                  (d) in {\sl Recent Progress in Many-Body Theories},
                  edited by T.L. Ainsworth, C.E. Campbell, 
                  B.E. Clements, and E. Krotscheck (Plenum, New York, 
                  1992), Vol. {\bf 3} p. 117; (e) {\sl J. Phys.:
                  Condens. Matter} {\bf 4}, 5783 (1992).

\bibitem{new18}  F.E. Harris,
                  {\sl Phys. Rev. B} {\bf 47}, 7903 (1993).

\bibitem{new19}   F. Cornu, Th. Jolicoeur, and J.C. Le Guillou,
                  {\sl Phys. Rev. B} {\bf 49}, 9548 (1994).

\bibitem{new20} R.F. Bishop, R.G. Hale, and Y. Xian,
                  {\sl Phys. Rev. Lett.} {\bf 73}, 3157 (1994);
                {\sl Int. J. Quantum Chem.} {\bf 57}, 
                919 (1996). 

\bibitem{new21} C. Zeng, I. Staples, and R.F. Bishop,
                {\sl Phys. Rev. B} {\bf 53}, 9168 (1996).

\bibitem{new22} R.F. Bishop, D.J.J. Farnell, and J.B. Parkinson, 
                {\sl J. Phys.: Condens.\ Matter} {\bf 8}, 11153 (1996).

\bibitem{new23} R.F. Bishop, Y. Xian, and C. Zeng, in 
                {\sl Condensed Matter Theories},
                edited by E.V. Lude\~na, P. Vashishta,
                and R.F. Bishop (Nova Science Publ., 
                Commack, New York, 1996), Vol. {\bf 11} p. 91.

\bibitem{new24} R.F. Bishop, J.B. Parkinson, and Y. Xian, 
                  {\sl Phys. Rev. B} {\bf 46}, 880 (1992).

\bibitem{new25} R.F. Bishop, J.B. Parkinson, and Y. Xian, 
                  {\sl J. Phys.: Condens. Matter} {\bf 5}, 9169 (1993).

\bibitem{ref7} D.J.J. Farnell and J.B. Parkinson, 
  {\sl J. Phys.: Condens.\ Matter}
  {\bf 6}, 5521 (1994).

\bibitem{ref8} Y. Xian, {\sl J. Phys.: Condens.\ Matter} 
  {\bf 6}, 5965 (1994).

\bibitem{ref9} R. Bursill, G.A. Gehring, D.J.J. Farnell, J.B. Parkinson, T. 
  Xiang, and C. Zeng, {\sl J. Phys.: Condens. Matter}\ {\bf 7}, 8605 (1995).

\bibitem{ref4}  R.R.P. Singh,
  {\sl Phys. Rev. B} {\bf 39}, 9760 (1989); 
  W. Zheng, J. Oitmaa, and C.J. Hamer, 
  {\it ibid.} {\bf 44}, 11869 (1991).                   

\bibitem{ref6} C.K. Majumdar and D.K. Ghosh,
  {\sl J. Math. Phys.} {\bf 10}, 1388 (1969);
  {\it ibid.} {\bf 10} 1399 (1969).
  
\bibitem{ref10} Chen Zeng and J.B. Parkinson, {\sl Phys. Rev.} {\bf B},
  11609 (1995).

\bibitem{ref11} J. H. Xu and C. S. Ting, {\sl Phys. Rev. B} {\bf 42},
  6861 (1990); A. V. Chubukov and Th. Jolicoeur, {\it ibid.} 
  {\bf 44}, 12050 (1991).
  
\bibitem{ref12} J. Richter, N.B. Ivanov and K. Retzlaff, {\it Europhysics 
    Letters} {\bf 25}, 545 (1994); 
  A. Voigt, J. Richter, N. B. Ivanov, 
  {\sl Physica A} {\bf 245}, 269 (1997).

\end{thebibliography}
\end{document}